# Investigation of Unit-1 Nuclear Reactor of the Fukushima Daiichi by Cosmic Muon Radiography

September 26, 2019

Hirofumi Fujii[1], Kazuhiko Hara[2], Kohei Hayashi[1], Hidekazu Kakuno[3], Hideyo Kodama[1], Kanetada Nagamine[1], Kotaro Sato[1], Shin-Hong Kim[2], Atsuto Suzuki[1$], Takayuki Sumiyoshi[3], Kazuki Takahashi[2#], Fumihiko Takasaki[1], Shuji Tanaka[1] and Satoru Yamashita[4]

[1] High Energy Accelerator Research Organization (KEK), Tsukuba, Ibaraki 305-0801, Japan
[2] University of Tsukuba, Tsukuba. Ibaraki 305-8571, Japan
[3] Tokyo Metropolitan University, Japan
[4] University of Tokyo, Japan

[$] Present address: Iwate Prefectural University
[#] Present address: Toshiba Co., Ltd.

## Abstract

We have investigated the status of the nuclear fuel assemblies in Unit-1 reactor of the Fukushima Daiichi Nuclear Power plant by the method called Cosmic Muon Radiography. In this study, muon tracking detectors were placed outside of the reactor building. We succeeded in identifying the inner structure of the reactor complex such as the reactor containment vessel, pressure vessel, and other structures of the reactor building, through the concrete wall of the reactor building. We found that a large amount of fuel assemblies was missing in the original fuel loading zone inside the pressure vessel. It can be naturally interpreted that most of the nuclear fuel was melt and dropped down to the bottom of the pressure vessel or even below.

## 1. Introduction

The nuclear power reactors of the Fukushima-Daiichi were heavily damaged by the giant earthquake and subsequent Tsunami occurred in March 2011. Decommissioning of damaged reactors is taking place. However, to proceed it, it is necessary to acquire information about the status of the reactors. Because of high radiation inside and around the reactor buildings, it is not easy to have an access for close inspection of the reactor status. We proposed to use a technique of Cosmic Muon Radiography (CMR) [1], [2] to study the current situation of the reactor, specifically the status of the nuclear fuel assemblies. As for a validation of the method, we constructed a detector system and evaluated its performance by imaging the nuclear reactor of the JAPC at Tokai, Ibaraki, Japan, by the CMR. We successfully imaged the inner structure of the reactor by detecting comic muons penetrating through the reactor by the detector placed outside of the reactor building [3]. Encouraged by the successful results, and after a study of the on-site radiation level, we carried out a CMR imaging of Unit-1 reactor of the Fukushima-Daiichi from February 2014 to June 2015. The Unit-1 reactor was damaged with its top concrete walls blown away by hydrogen-gas explosion, as in a photograph of Fig, 1, provided by the TEPCO, the Tokyo Electric Power Company Ltd. During our investigation, the whole building was covered to protect any debris to fly out. The details of the investigation are given in this report.

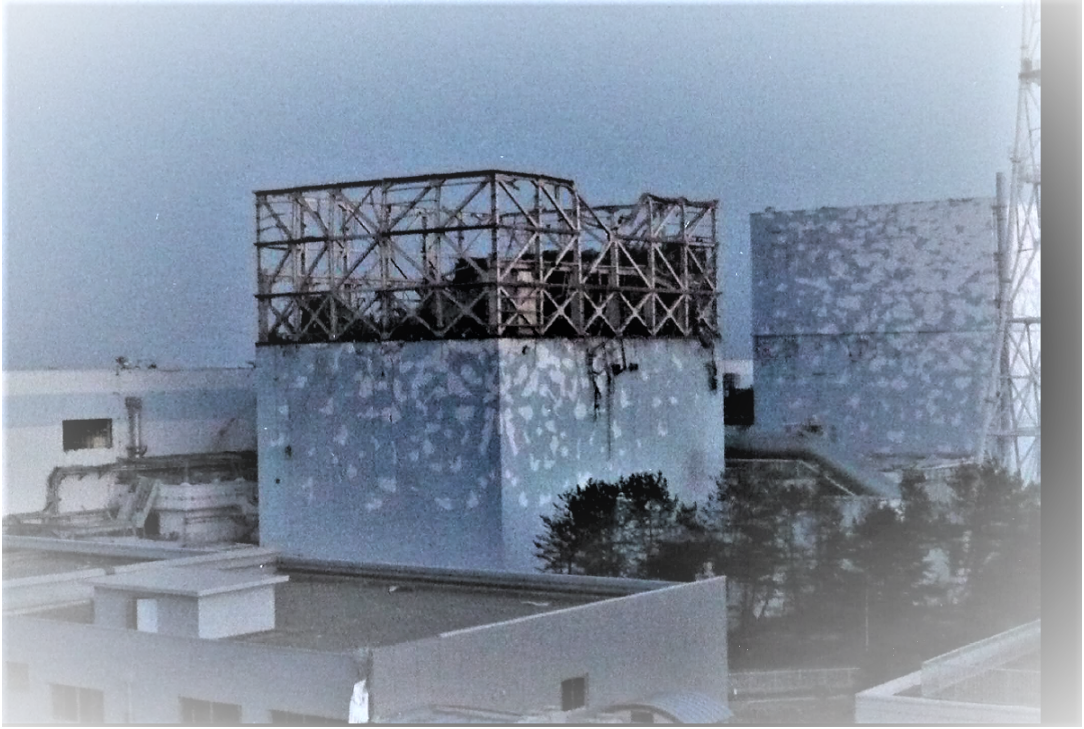

Fig. 1. Damaged Unit-1 Reactor of the Fukushima-Daiichi, viewed from north-west side. The Unit-2 building is seen to the right. (Courtesy of TEPCO).

## 2. The Muon Detection System

The detection of the cosmic muons was made by X-Y sets of 1-cm wide plastic scintillation bar counters (PSC) which are arranged in a plane of 1 m×1 m. While we used two X-Y sets of the PSC plane for the muon telescope system at the JAPC [3], we used three X-Y sets at the Fukushima-Daiichi reactor to be robust against the possible accidental coincidence due to the environmental high radiation. The muon telescope was housed in a container with 10-cm thick iron to suppress the effect of the environmental radiation.

*2.1 Protection of the detector against the environmental radiation*

The major concern in operating the detector on-site at the Fukushima-Daiichi was the effect of the environmental radiation, which was reported as high as 0.5 mSv/h around the Unit-1 reactor building. There were some "hot spots" in addition where the radiation level was much higher. The main components of the environmental radiation were mainly 662 keV gammas from $Cs^{137}$ and gammas (mainly 605 and 796 keV) from $Cs^{154}$, and their Compton electrons may deteriorate the cosmic muon tracking. We considered to shield the detector with iron, and to evaluate the necessary thickness of the iron shield, we performed a "test tracking" by a sample detector placed in a 5-cm thick iron box. The sample detector was made of three layers of ten PSCs, the same PSCs used for the muon telescope. The iron shield box was additionally surrounded with iron plates of 5- or 10-cm thicknesses, therefore we obtained data for iron thicknesses 5, 10 and 15 cm.

Figure 2 shows the sample detector placed in the 5-cm thick iron shield box. We brought this system near to the Unit-1 reactor building and studied the effect of the environmental radiation to cosmic muon tracking.

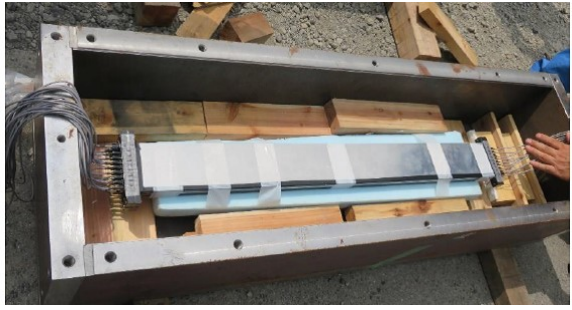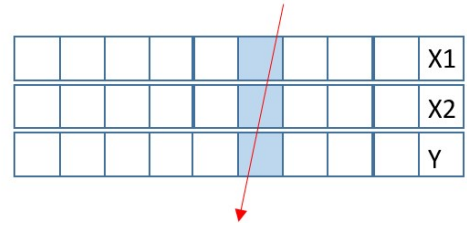

Fig 2: Sample detector (10 cm wide×1 m long×10 cm thick) placed in 5-cm thick iron shield box. (right) Cross-sectional view of the three-layer sample detector, indicating grouping of the three layers.

The suppression factors of the gamma rays from $Cs^{134}$ and $Cs^{137}$ were estimated by a GEANT simulation as a function the iron thickness assuming gamma rays to hit normally to the iron [4]. The gamma yield is expected to be suppressed to 0.3 with 5-cm thick iron and to 0.03 with 10-cm thick iron, as shown in Fig. 3. To compare with the simulation results, we first used a handy radiation monitor tightly enclosed in 10-cm thick iron plates to measure directly the suppression factor at the site planned for deployment of the detector. The measured suppression is close to of 15-cm simulation, as plotted in Fig. 3, as the difference can be explained by the average path length in the measured configuration was 13 cm taking into account that the gammas are coming from any directions.

The signals of the sample detector were read out by the same system [3] as for the muon telescope where existence of coincidence hits in two layers is required. For the three-layer sample detector, we assigned the two top layers as "X" and the bottom layer as "Y", requiring at least one hit in each group of "X" and "Y" (note all three scintillator bars run in the same direction despite for the naming X or Y). Figure 4 shows the "X1-Y" coincidence rates measured on site at the Fukushima-Daiichi for the three thicknesses of the iron shield as a function of the width of coincidence time window. As the accidental coincidence increases linearly with the coincidence time window width, the measured rate is affected by the environment radiation for the time window longer than 100 ns for 5-cm iron, and started to be affected above 1000 ns for 10-cm iron.

The typical time window we employ for the muon telescope is 16 or 32 ns. This time window needs to be multiplied by a factor 5 to 10 in evaluating the effect in the complete telescope using the present results of the sample detector, since the number of the scintillator bars (100 in the complete system) to be considered is more by this factor. We conclude that the coincidence rate is barely affected for the iron shield thicker than 10cm.

Figure 5 shows the hit position correlation between "X1" and "Y" layers when exactly one hit is required to be found in each of "X1" and "Y". As expected, the correlation is clearly seen for the 10-cm iron up to 1024 ns time window. For the 5-cm iron, the correlation is less clear and influenced more with increasing the time window width. The study was repeated for relaxed threshold setting, 0.8 times the nominal setting, to investigate the effect of the possible uncertainty in the threshold determination, and found small dependence.

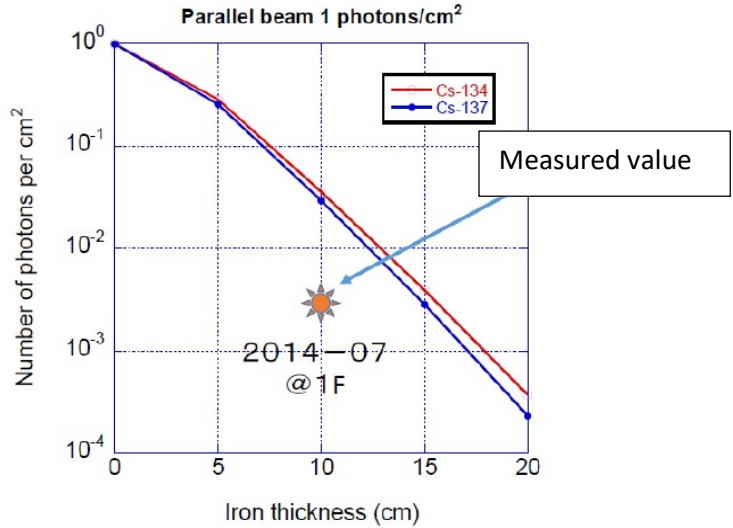

Fig. 3. Suppression factor of the gamma rays from $Cs^{134}$ and $Cs^{137}$ estimated by a GEANT simulation as a function of the thickness of the iron, assuming gamma rays are injected normally to the iron. The value measured on site for 10-cm iron is shown by star.

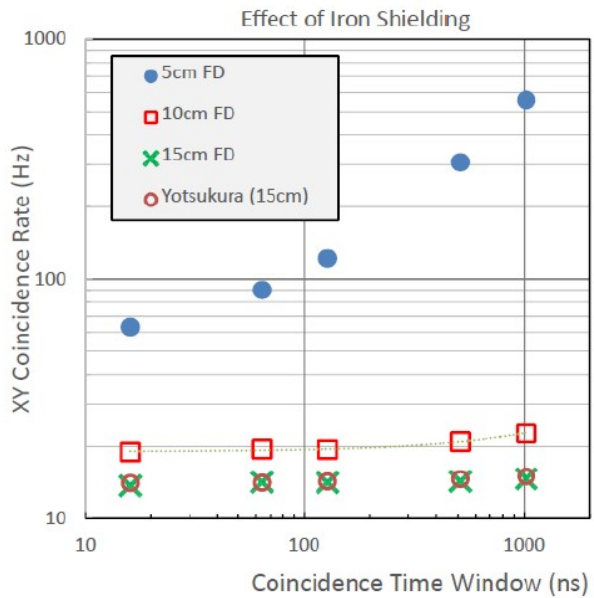

Fig. 4. Coincidence rates measured with the test detector as a function of the time window of the coincidence. The data are shown for the three iron thicknesses measured at the Fukushima-Daiichi site, and one measured at Yotsukura, about 50 km south. The ambient radiation level at Yotsukura was below one µSv/h.

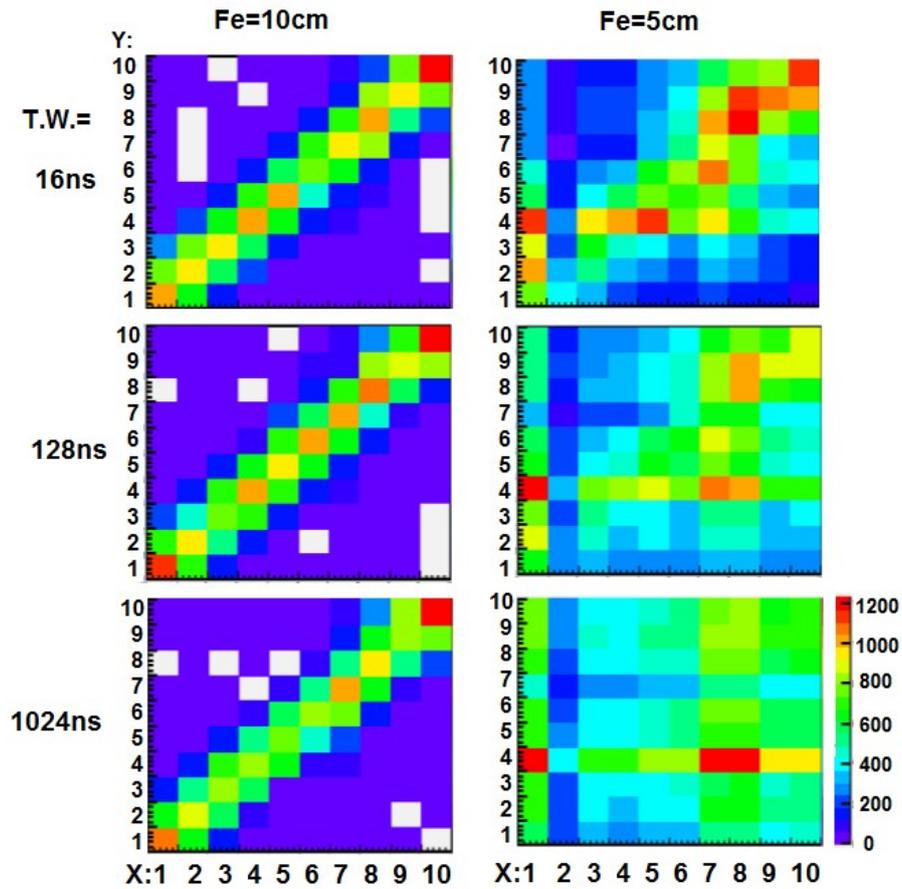

Fig. 5. X2 and Y hit correlation measured with the test detector; (left) for 10-cm iron and (right) for 5-cm iron shield. The results are shown for three typical time window widths, 16, 128 and 1024 ns.

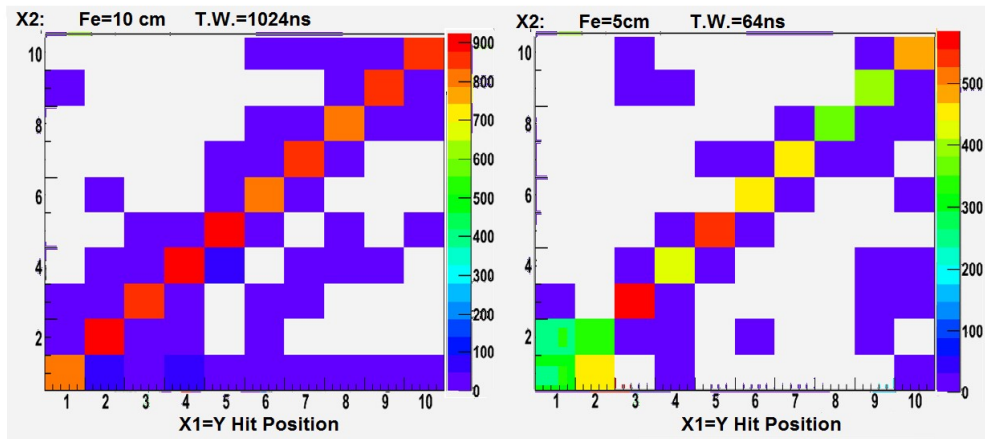

Fig. 6. X2 and Y=X1 hit correlation measured with the test detector; (left) for 10-cm iron and (right) for 5-cm iron shield. The time window widths are 1024 ns and 64 ns, respectively. The number of events in each column sum is normalized for N(Y=X1) =1000.

Finally we examined whether we can find a hit in the middle layer X2 when X1 and Y recorded hits at the same channel number. The hit correlations are shown in Fig. 6 for typical cases with 10 and 5-cm iron shield thicknesses. The plots are normalized such that the number of X1=Y events is 1000. X2 hit can be found at the proper position with efficiencies more than 90% for 10-cm iron case with time window of 1024 ns. The observed efficiency is not large for 5-cm iron, although the correlation itself is not much distorted. This can be explained by that the reference Y=X1 is spurious at some fraction and the corresponding X2 was not found as the event is spurious: once the reference is correct, X2 is found correctly.

2.2 The detector system for Unit-1 Observation

Based on the test measurement data and further considerations, we concluded that there will not be any problem in tracking with 10-cm iron. With 5-cm iron, X-Y hit correlation is distorted in a single X-Y assembly. With 5-cm iron and two X-Y assemblies combined, the signal-to-noise ratio will be improved to about 100 for the nominal threshold setting, however, limited to 5 for relaxed threshold setting (0.8 times the nominal).

For the detector system to be deployed for the Unit-1 observation, we selected an iron shield thickness of 10 cm and chose conservatively to install three sets of X-Y assemblies. Figure 7 shows one of the Muon Telescope systems. The detection system was equipped with an air-conditioner to provide a constant temperature inside the iron shield to be kept 20±3℃ for a stable operation of the photo-device, multi-pixel photon counter (MPPC).

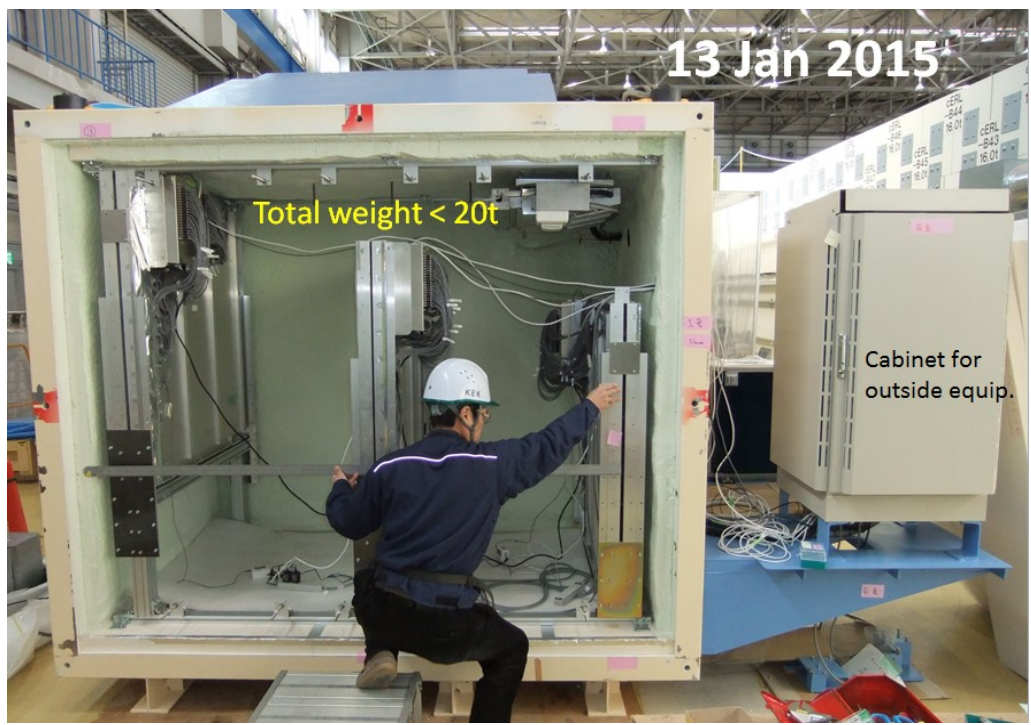

Fig. 7. The muon telescope housed in an iron shield box made of 10-cm iron plates. The relative elevation heights of the three X-Y assemblies are adjusted to target at the Unit-1 fuel loading zone. For thermal insulation, 6-cm thick plastic foam is plastered over the inner wall of the box. The cabinet next to the box houses the air-conditioner controller and data network hub for data transfer and communication with the assemblies. The total weight is 20 t.

The X-Y assemblies used for the study at the JAPC at Tokai were refurbished repairing dead channels and two Muon Telescope systems were constructed at KEK. They were moved to the Fukushima-Daiichi site, and placed in January 2015 at locations marked by red circles in Fig. 8 which shows a plane view of the Unit-1 reactor. The two systems were placed at WN and NW initially, and one of them was moved to the point N later in summer. The distance to the reactor center is 36 m for the location WN.

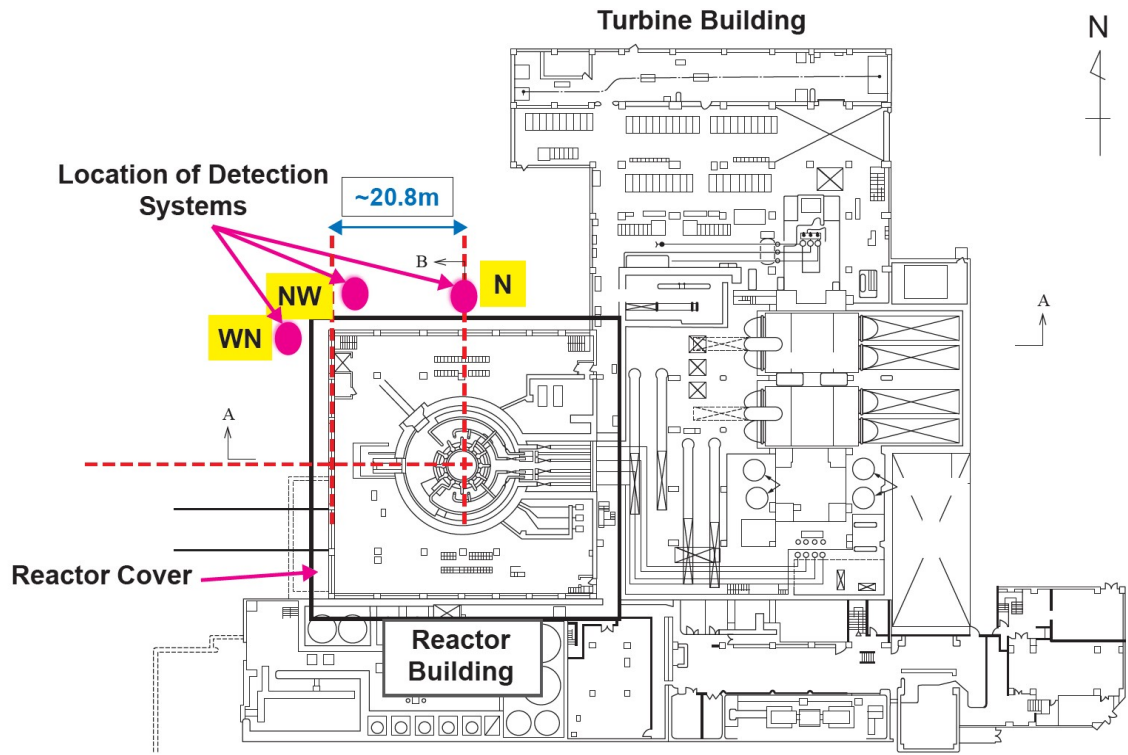

Fig. 8. A plane view of the Unit-1 reactor of the Fukushima-Daiichi. The observation of the reactor inner structure was made from the three locations marked by red circles. (Courtesy of the TEPCO).

*2.3 Images of the Unit-1 reactor*

The observation of the cosmic rays started in February 2015. We confirmed the detection systems were functioning properly in an environment of the radiation level of about 500 μSv/h. Figure 9 shows the Unit-1 reactor images after 90 days, representing the lateral muon absorption distribution, the observed event rates normalized by those obtained at KEK without obstacles in front. Larger absorption due to heavy objects is illustrated in darkness.

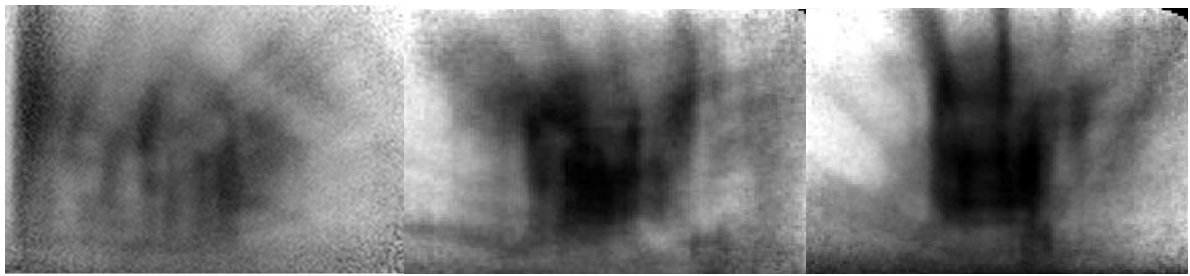

Fig. 9. Images of the Unit-1 reactor of the Fukushima-Daiichi after 90 days observed by the muon telescope placed at (left) WN, (middle) NW and (right) N points given in the figure 8.

The figure 10 illustrates the observed image by the muon telescope placed at the WN point with the Reactor Pressure Vessel marked with the red-dot line. The figure 11 shows the same image with the identification of the facilities in the reactor building. One notices that no sign of the existence of heavy materials such as the lump of nuclear fuel in the RPV. One naturally interprets that the nuclear fuel got melted and dropped from the original position in the RPV.

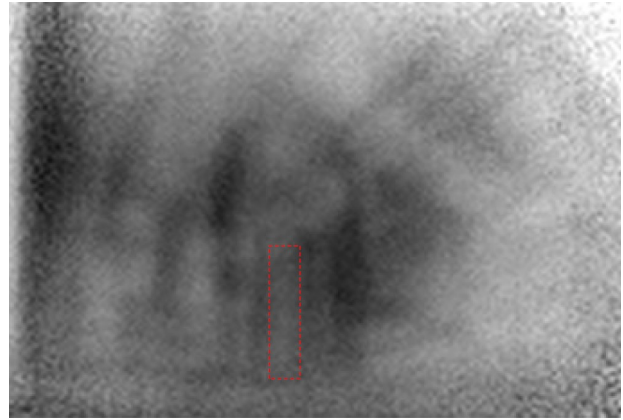

Fig. 10. The observed image from the WN point with the Reactor pressure Vessel marked by the red-dot line

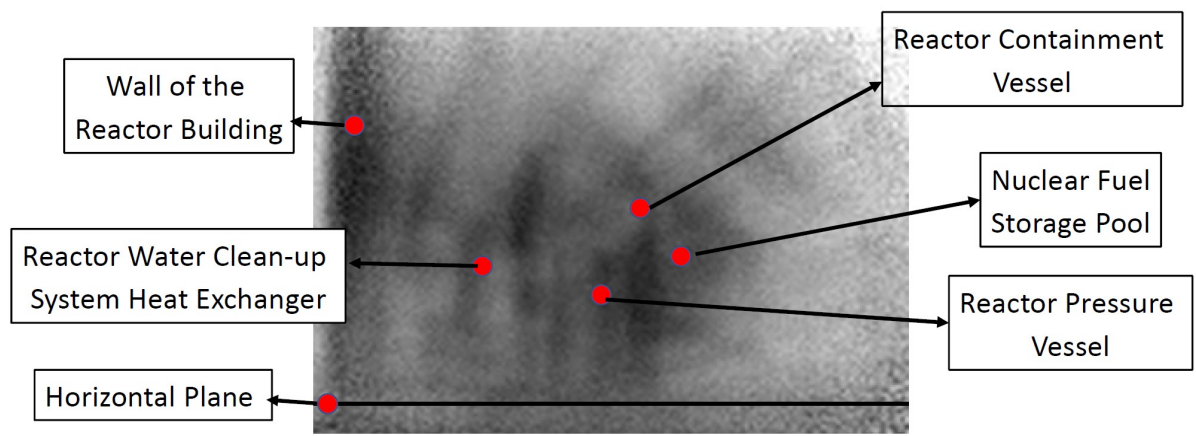

Fig. 11. Image of the Unit-1 reactor of the Fukushima-Daiichi after 90 days observed by the detection system placed northern west corner of the reactor building.

The distribution pattern of the muon flux reduction can be translated to the distribution of heavy structures inside the reactor complex, as shown in Fig. 10 for the muon telescope placed at the northern west corner (marked as WN in Fig. 8) of the reactor building. We identified the structures such as 1) the reactor pressure vessel (RPV), 2) the reactor containment vessel (RCV), 3) the Nuclear Fuel Storage Pool, and 4) the north wall of the reactor building. From the three images, we concluded that any heavy lump of materials is not seen at the core part of the pressure vessel. This observation can be interpreted that most of the loaded nuclear fuel assemblies melted and dropped from the loading zone.

3. Estimation of the amount of the nuclear fuel left in the RPV of the Unit-1 reactor

The amount of heavy materials left in the RPV is a crucial information to be investigated with the present detection method. Figure 12 illustrates the structure of the Unit-1 reactor. We defined "side-band" regions and evaluated the amount of material in the "core" region from the amounts of material estimated in the side-band regions. We defined three vertical slice regions, (a), (b1) and (b2) by the point-dash vertical lines shown in Fig. 12.

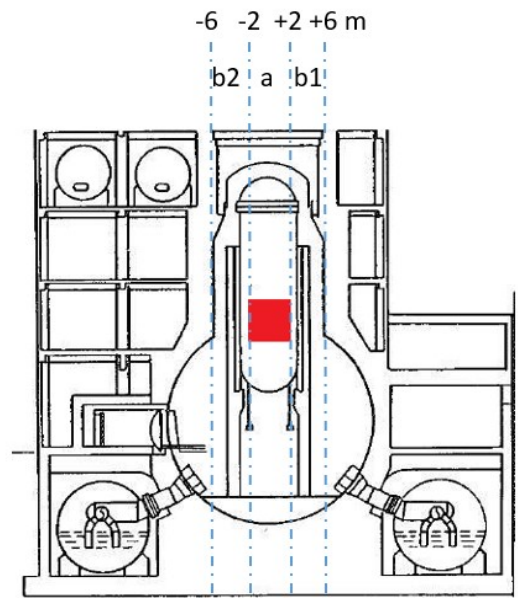

Fig 12. Structure of the Unit-1 Nuclear Reactor of the Fukushima-Daiichi. The red box represents the fuel loading zone.

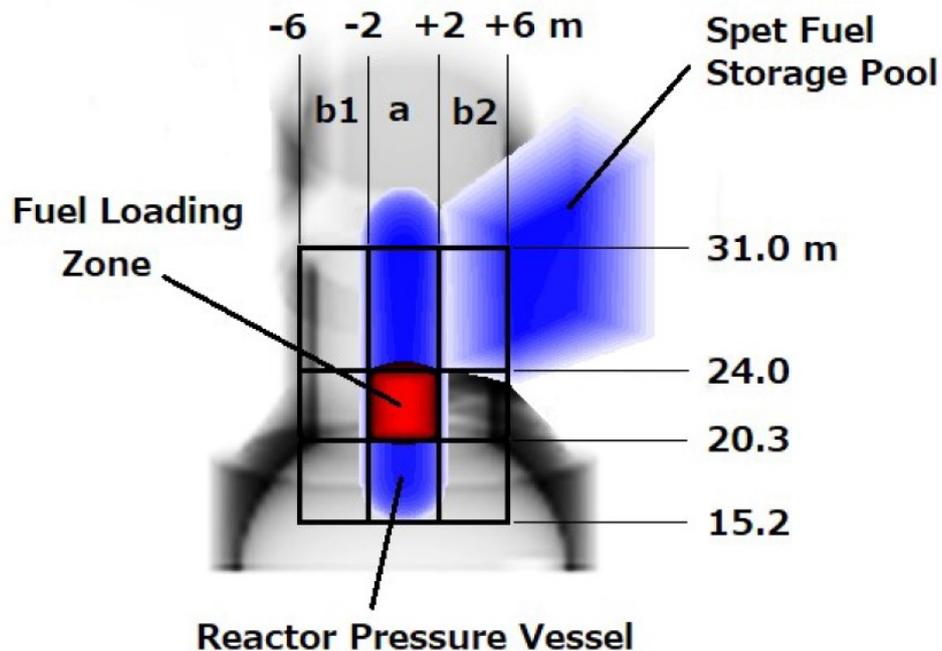

Fig 13. An artistic view of the reactor complex of the Unit-1 reactor with the Reactor Pressure Vessel and the Spent Fuel Storage Pool

The amount of materials through which cosmic muons go through is expressed by the density length which is defined as the density of materials traversed by the muon multiplied by the muon path length. Specifically, for a given zenith angle θ, the muon momentum distribution function $D(\theta:p)$ is used such that the threshold momentum $p_0$ corresponds to the observed muon below which muons cannot go through materials to be observed. Namely, we solve the following equation:

$$\int_{p=p_0} D(\theta:p)dp \Big/ \int_{p=1} D(\theta:p)dp = N_{obs}/N_0 ,$$

where $N_{obs}$ and $N_0$ are number of detected muons and number of expected with no material in front (KEK measurement), respectively, and we set the minimum momentum to $p$=1 GeV. Then the threshold momentum $p_0$ is translated to the material path length from the muon range, where we assume that the material is concrete of 2.5 g/cc density. In the calculation the cosmic muon momentum distribution $D(\theta:p)$ is taken from the theoretical calculation by Smith-Duller [5] or from a measurement by Okayama group [6]. The difference between two assumptions is assigned as a systematic uncertainty. More detailed description of the present method is given in Appendix 1.

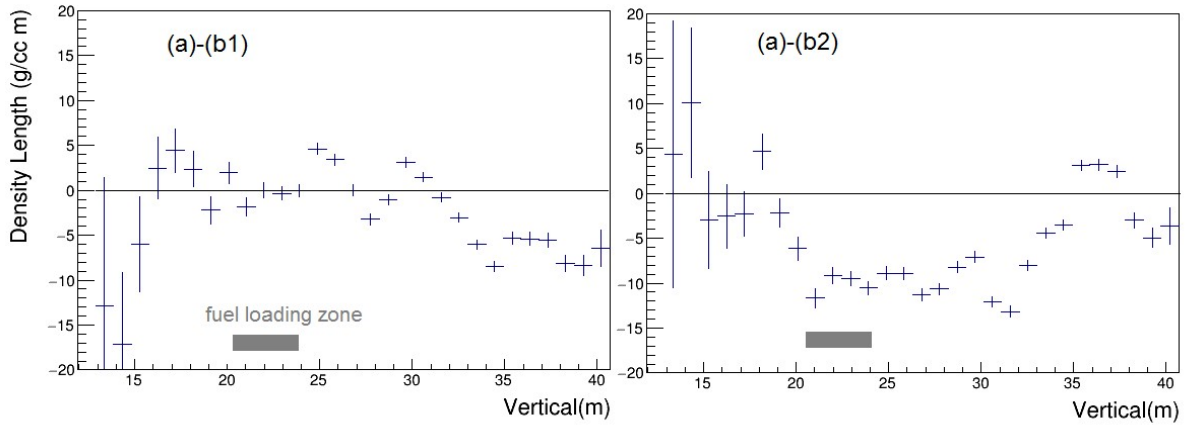

Fig 14. (left) The density length along the slice (b1) subtracted from that along slice (a); (right) the density length along the slice (b2) subtracted from that along slice (a). The vertical range corresponding to the fuel loading zone is indicated by the shaded box in the figures.

The difference of the obtained density-length distribution (a)-(b1) and of (a)-(b2) are plotted in Fig. 14. Although the material distributions in these slices are not identical, possible zenith angle dependence of the muon flux can be diminished by subtraction. In practice, the density length along (b2) is much larger due to the nuclear fuel storage pool located behind the slice of (b2). Therefore, we take only the difference (a)-(b1) for evaluating the amount of material remaining in the fuel loading zone.

The evaluated amounts in terms of ton are summarized in Table 1. The subtraction results are shown separately for different height ranges, one covering the fuel loading zone, others underneath and above the loading zone. We independently calculated the weights in the corresponding regions from the engineering drawing of the Unit-1 reactor and obtained 72 tons for the (a)-(b1) weight in the region above the loading zone, while we measured 32±6(stat)±11(syst) tons. The difference is taken as a systematic uncertainty and added conservatively to the amount by taking the different height coverage into account. The amounts shown in the row (a) of Table 1 are resulting values corresponding to "excess" in the slice (a). The systematics uncertainty is composed of the material difference between the measurement and calculation (described above), difference of the momentum modelling and the difference of the detector efficiency measured at KEK and on-site at the Unit-1 reactor. Among these, the momentum modelling is largest. It is larger in the height range underneath the loading zone because available muon flux data is limited in this zenith angle range.

Table 1. Measured amount of material in the RPV for different height regions.

| region | 15.2 m < h < 20.3 m (beneath loading zone) | 20.3 m < h < 24.0 m (loading zone) | 24.0 m < h < 31m (above loading zone) |
| --- | --- | --- | --- |
| (a)-(b1) | 4±25(stat)±32(syst) tons | 1±7(stat)±5(syst) tons | 32±6(stat)±11(syst) tons |
| (a) | 33±25(stat)±43(syst) tons | 22±7(stat)±22(syst) tons | 72(*) |

(*: calculated from the engineering drawing)

The amount of the fuel assemblies originally at the loading zone is estimated to be 160 tons. From the observation described above, we come to a conclusion that most of the nuclear fuel loaded initially at the loading zone in the RPV melted and dropped to the bottom of part of the RPV or even below penetrating through to the RCV.

Unfortunately, we cannot further detail the status of the fuel debris around the bottom part of the RPV because the cosmic muon flux is limited in the present arrangement of the muon telescope located at the corner of the reactor building. At the location N shown in Fig. 8, the fuel storage pool behind the PRV degrades the measurement. It would be necessary to lower the telescope and set it closer to the RPV so that the bottom part of the reactor can be viewed.

## 5. Summary

We studied the fuel stauts of the Unit-1 reactor of the Fukushima-Daiichi using the cosmic muon telescope detectors placed outside the reactor building. We succeeded in identfying major stuctures of the reactor complex, reactor containment vessel, pressure vessel, and others. The obseved absorption map clearly indicates that heavy material causing significant attenuation of cosmic muons is missing in the fuel loading zone inside the pressure vessel. We naturally interpret that the nuclear fuel assemblies had melted and dropped to the bottom of the pressure vessel or even below. The amounts of meterial in the fuel loading zone and below were estimated and the results are consistent with the above conclustion.

We tried to detail the bottom part of the pressure vessel in search for strong attenuation of the cosmic muons, however, no evidence of strong attenuatin was identified. To understand this result, two possibilities are considered: 1) the nuclear fuel melted and dropped beyond the scope of the prensent detector coverage, and 2) the energy of cosmic muons at sharrow angles is so high that they penetrate the nuclear debris possible at the bottom part of the RPV. We are currently studying if the cosmic rays at shallow elevation angles can penetrate large amount of materials, therefore we can not learn the structure of the nuclear reactor in the corresponding angle range by the method employed in this study.

**Acknowledgements**

The present study has been carried out as a project supported by the International Research Institute for Nuclear Decommissioning IRID. The cooperation and lots of help provided by the TEPCO and Tokyo Power Technology Ltd. have been inevitable for the detector installation and operatoin. We thank Professor Hirotaka Sugawara of the Okinawa Institute of Science and Technology for many advices and suggestions.

**Appendix: Derivation of the effective mass of the fuel debris**

We describe the calculation methods employed to dervie the mass of the object from the observed muon transmission. In overall, the muon transmission was defined as the ratio of the flux distributions measured with and without the object in front. The data without the object were collected at KEK or other places where no object was located in front. This method reduces the uncertainty originating from the detector non-uniformity, for example.

The nuclear plant building is composed mainly of shielding concrete. Figure A-1 illustrates the density-length (density multiplied by the path length) distribution of the Unit-1 nucler plant viewd from the NW point, which was calculated using the available design drawings. It distributes in the range 30-70 m·g/cm$^3$ vertically along the axis of the pressure vessel, the target region to be investigated. The density-length range in other views is within 20-80 m·g/cm$^3$ and the corresponding minimum muon momenta to pass through are 4-17 GeV/c, calculated based on the continuous-slowing-down approximation (CSDA) range, see Fig. A-2.

The view area of the detector, as shown in Fig. A-1, is wide enough covering the pressure vessel (area to be investigated) and other reactor building structures. The structures can be assumed to be composed of concrete. Therefore the flux data outside the area to be investigated are related to the density-length distribution calculated using the design drawings, as a function of the muon zenith angle. Since the muon momentum is strongly dependent on the zenith angle, this mothod provides more reliable estimate than simply relying on the muon momentum distributon projected from the available flux measurement. Also in the calculatoin using the projected muon momentum distribution, we would need to assume the flux reduction to be caused by the CSDA range.

In the search area, we need to consider any CSDA range difference between the concrete and other materials. Figure A-3 plots the calculated CSDA range values as a function of the muon momentum for the typical materials in the search area. It should be noted that the absolute values are not identical but the momentum dependences are similar within ±5%. The muon flux reduction should be a function of *a*ρ*L* where *a* is the absorption coefficient related to the CSDA range, ρ is the density and *L* is the path length. The difference in the absolute values is taken into the systematic uncertainty in *a*.

The density-length distribution derived from the observed muon flux reduction map are subtracted with the calculated density-length distribution of the structures in the front and in the rear of the search region, namely the pressure vessel. Or the subtraction was conducted using the region next to the search region. The remaining density-length distribution was multiplied with the size of search area to derive the amount of material left in the presseure vessel.

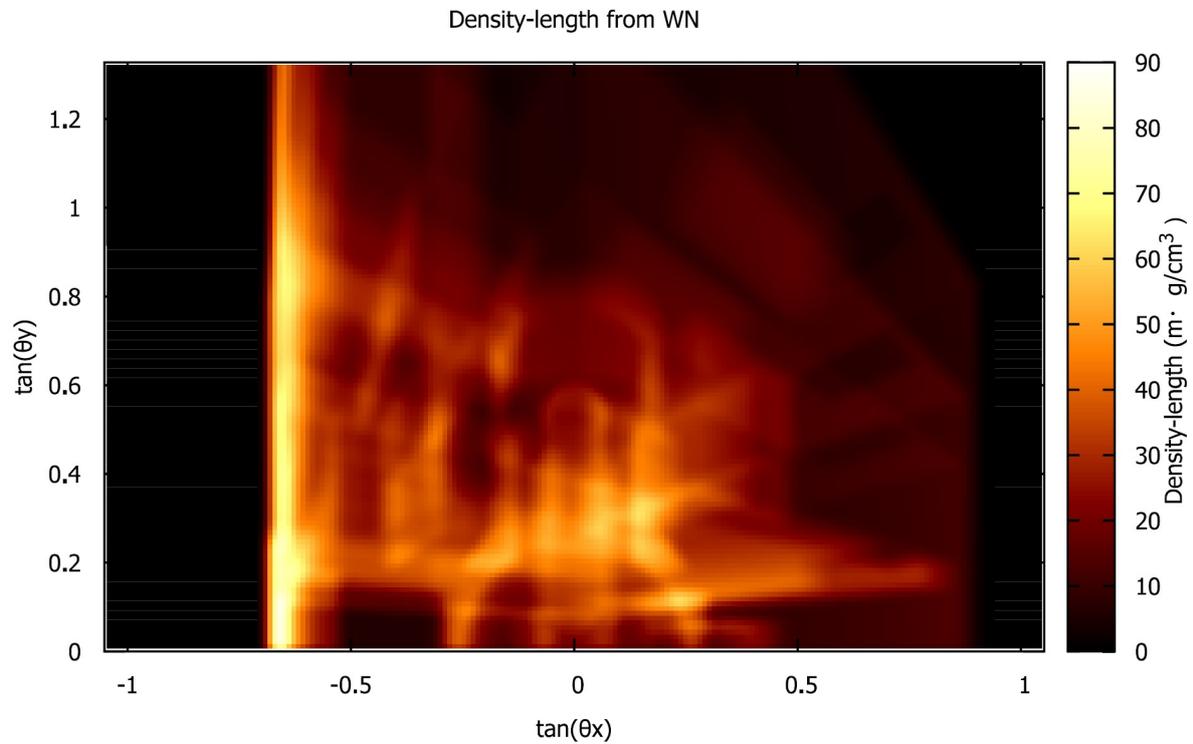

Fig. A-1 Density-length distribution (in m·g/cm3) viewed from the WN point in the plane of detector view area. The axes are horizontal and vertical directions of the muon tracks.

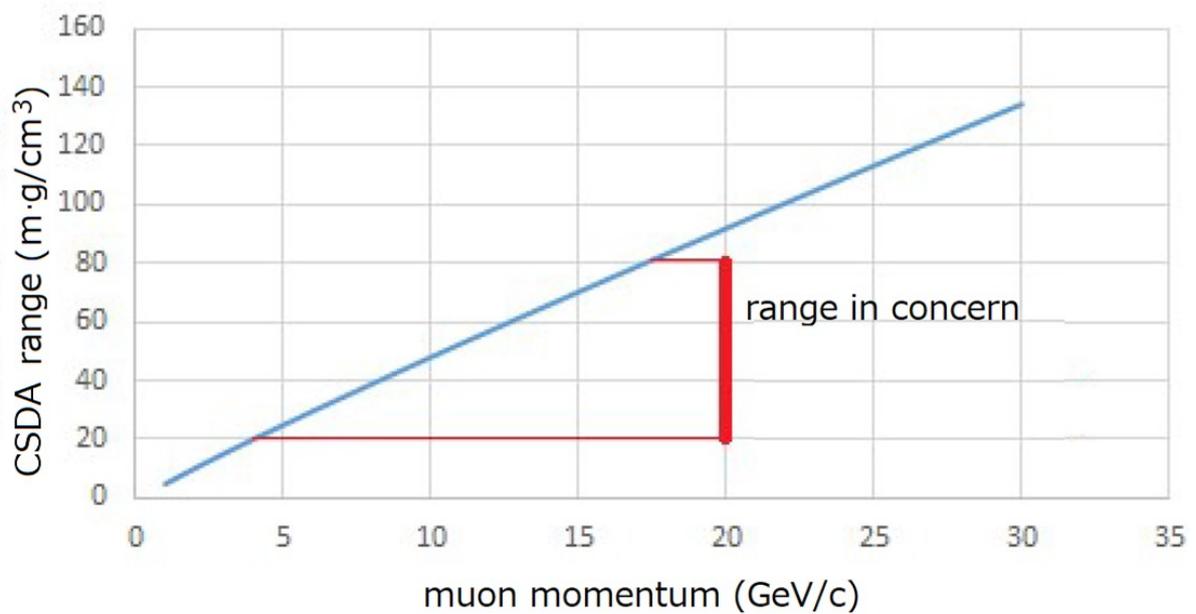

Fig. A-2 Correlation between the minimum muon momentum and the muon range. Assumed material is concrete of standard density 2.3 g/cm³.

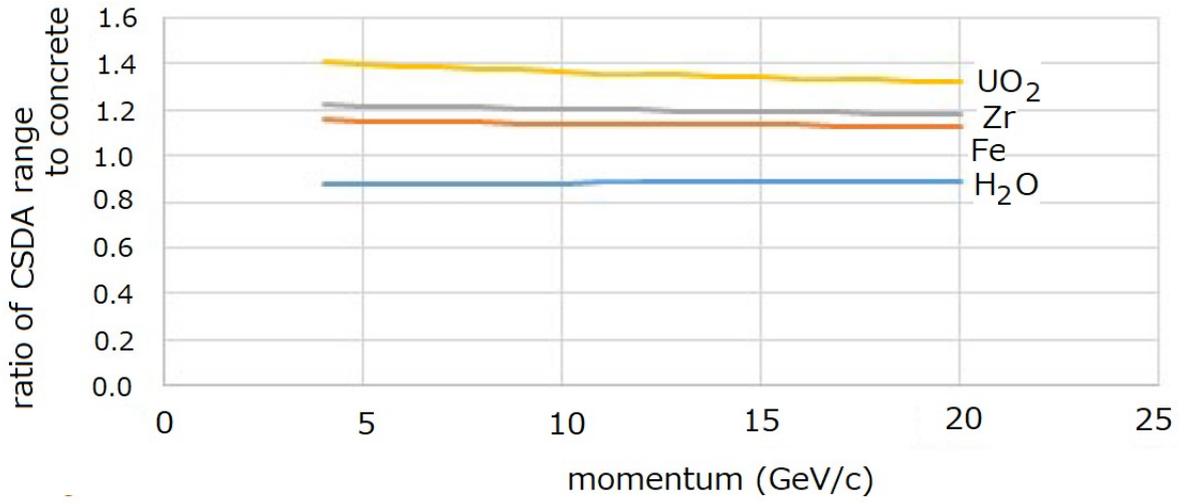

Fig. A-3 Calculated ratio of CSDA range to concrete for the materials abundant in the pressure vessel as a function of the muon momentum. The ratio for $UO_2$ varies from 1.41 to 1.32 (a change of 10%) in the momentum range in concern.